\documentclass[a4paper,twocolumn,unpublished]{quantumarticle}
\pdfoutput=1
\usepackage[utf8]{inputenc}
\usepackage[T1]{fontenc}
\usepackage[english]{babel}
\usepackage{amsmath}
\usepackage{amssymb}
\allowdisplaybreaks

\usepackage{xcolor}
\usepackage{graphicx}
\usepackage{float}
\usepackage{bm}

\usepackage[italicdiff]{physics}
\usepackage[numbers,sort&compress]{natbib}
\usepackage[colorlinks=true,allcolors=quantumviolet]{hyperref}
\usepackage{microtype}

\begin{document}

\title{Gravitational Entanglement in Optomechanics:\newline Distinguishing Classical and Quantum Models}

\author[1]{Samuel Schlegel}
\orcid{0009-0005-3248-3208}

\author[2,3]{Ankit Kumar}
\orcid{0000-0003-3639-6468}

\author[4,5,6]{Tomasz Paterek}
\orcid{0000-0002-8490-3156}

\author[1,7]{Borivoje Daki\'c}
\orcid{0000-0002-8490-3156}

\affil[1]{University of Vienna, Faculty of Physics, Vienna Center for Quantum Science and Technology, Boltzmanngasse 5, Vienna 1090, Austria}

\affil[2]{Schulich Faculty of Chemistry, Technion – Israel Institute of Technology, Haifa 3200003, Israel}

\affil[3]{Helen Diller Quantum Center, Technion – Israel Institute of Technology, Haifa 3200003, Israel}

\affil[4]{School of Mathematics and Physics, Xiamen University Malaysia, 43900 Sepang, Malaysia}

\affil[5]{Institute of Theoretical Physics and Astrophysics, Faculty of Mathematics, Physics and Informatics, University of Gdańsk, 80-308 Gda\'{n}sk, Poland}

\affil[6]{Centre for Photonics and Quantum Communication Technology, Indian Institute of Technology Roorkee, Roorkee 247667, India}

\affil[7]{Institute for Quantum Optics and Quantum Information, Austrian Academy of Sciences, Boltzmanngasse 3, Vienna 1090, Austria}

\maketitle

\begin{abstract}
Observation of gravitationally induced quantum entanglement is often interpreted as a direct evidence of non-classical gravity. While the form and the degree of non-classicality have been rigorously studied from a foundational perspective, classical models reproducing experimental signatures of such entanglement remain underexplored. Motivated by the experimental simplicity, nearly all existing optomechanical approaches assume Gaussian initial states, and due to the weakness of gravity the quantum Newtonian potential is truncated at the second order. However, this regime admits a classical description in terms of the Wigner-Weyl representation, including features typically associated with quantum entanglement. A clear distinction between classical and quantum predictions emerges only beyond this setting. We comprehensively analyze the possibilities and provide operational witnesses for detection of non-classicality via Wigner negativity, and detection of non-quantumness via negativity of the Weyl operator. Our results demonstrate that the experimental requirements on certifying gravitational entanglement are more stringent than previously anticipated.
\end{abstract}

\section{Introduction}

The idea that two masses could be entangled solely through their gravitational interaction dates back to Feynman~\cite{DeWittRickles2011}.
Concrete proposals for realizing such experiments in the near future have emerged only recently~\cite{Bose2017,Marletto2017,Krisnanda2020}, and their foundational implications remain under active debate.
Different models of the mass---field---mass system attribute gravitationally induced entanglement to different forms of non-classicality.
For example, it has been linked to non-classical correlations with the mediator~\cite{Krisnanda2017}, non-commuting field observables~\cite{Marletto2017}, non-zero off-diagonal terms in coherent state basis~\cite{Bose2017}, and, more recently, to superpositions of geometries~\cite{Chen2023} and quantum features of spacetime~\cite{Christodoulou2023}, and non-commuting field-particle couplings~\cite{Ganardi2024}.
At the same time, hybrid models combining classical gravity with quantum matter~\cite{Diosi1987,Penrose1996,Tilloy2016,Oppenheim2023} can also produce entanglement~\cite{trillo_diosi-penrose_2025}, leading to the view that only detailed measurements can rule out such alternatives~\cite{Aziz2025,DiBiagio2025}.

Here, we set aside these foundational questions and instead focus on a practical reason why the mere detection of certain signatures of gravitational entanglement is not sufficient to exclude classical gravity.
A natural route towards observation of this form of correlation is to leverage the methods of optomechanics~\cite{Aspelmeyer2014}, which enable preparation of near-ground Gaussian states in harmonic potentials and precise quadrature measurements \cite{Chan2011}.
Since gravity is weak, the relative displacement of the masses in such experiments is much smaller than their separation, and hence existing proposals typically rely on a second-order expansion of the quantum Newtonian potential in this small parameter.
Although this approximation already predicts entanglement, in fact all signatures accessible through quadrature measurements in this regime also admit a classical explanation.
This classical model is just the Newtonian evolution of phase-space distributions, obtained via the Wigner-Weyl transformation.
This does not contradict the presence of quantum signatures, as the corresponding classical model involves states which would violate the Heisenberg uncertainty relation.

To certify a genuinely quantum behavior, some of these assumptions must be relaxed. We comprehensively analyze two such routes: the preparation of non-classical initial states of the masses, and the inclusion of the third-order term in the gravitational potential. 
We also discuss additional possibilities, such as measurements beyond standard quadrature observables and the use of independent evidence that the individual masses themselves exhibit quantum behaviour.
In all of these scenarios, classical and quantum predictions diverge. In particular, incorporating the cubic contribution to the quantum gravitational interaction generates both entanglement and negativity of the Wigner function, providing a direct signature of non-classicality. By contrast, the corresponding classical model retains a positive Wigner function while still reproducing entanglement signatures. 
However, the associated Weyl operator acquires negative eigenvalues, thereby signaling non-quantumness.
We show how these features can be detected experimentally. 
Finally, we note that a systematic analysis of various notions of entanglement within the Wigner–Weyl framework was recently presented in Ref.~\cite{Schlegel2025}.
As we show below, gravitational dynamics provides a natural setting in which all of these notions can arise.

\section{Scenario and prerequisites}

From an experimental point of view, the simplest setup for observing gravitational entanglement consists of two identical spherical masses $m$ constrained to move along one dimension $x$, with initial center-to-center separation $L$. 
The masses are independently prepared in suitable initial states, allowed to evolve under purely gravitational interaction, and finally measured using quadrature apparatus.
This scenario is particularly interesting because it admits multiple modeling approaches, and the resulting measurements can be used to distinguish between them. Below, we compare predictions from quantum and classical mechanics and show that they overlap significantly under typical assumptions.

We begin with the quantum description. Consider two identical masses prepared in a product of identical Gaussian states of width $\sigma$:
\begin{align}
\psi_j(x_j) & =  \frac{1}{ \qty( 2 \pi \sigma^2 )^{1/4} } \exp \qty( - \frac{x_j^2}{4 \sigma^2}  ),
\quad j = 1,2,
\label{EQ_ISTATE}
\end{align}
where $x_1$ and $x_2$ are the respective displacements of the two masses from their initial average positions, and for simplicity we assumed the initial relative momentum as zero. 
In this notation, the Hamiltonian reads:
\begin{align}
H = - \frac{\hbar^2}{2 m} \partial^2_{x_1} - \frac{\hbar^2}{2 m} \partial^2_{x_2} - \frac{G m^2}{L + x_2 - x_1}.
\end{align}
For simplification we make a transformation to the center of mass (COM) frame of reference, where $R = (x_1+x_2)/2$, and $r=x_2-x_1$,
are the displacements of the center of mass and the reduced mass, respectively.
In these coordinates the Hamiltonian is separable in the two degress of freedom, $H = H_R + H_r$, where 
$H_R = - \frac{\hbar^2}{4 m} \partial^2_{R} $  and
\begin{align}
H_r &=  - \frac{\hbar^2}{m} \partial^2_{r} + V(r) ,
\quad
V(r) = - \frac{G m^2}{L + r} .
\end{align}
The initial state remains a product of Gaussians $\psi_1(x_1) \psi_2(x_2) = \psi_R(R) \psi_r(r)$, where the position spreads of the COM and the reduced mass are $\sigma / \sqrt{2}$ and $\sigma \sqrt{2}$, respectively.
Given the separability, the COM wave function $\psi_R(R)$ evolves freely, and hence the problem reduces to evolving the wavefunction of the relative motion $\psi_r
(r)$ under the potential that depends only on the relative distance.

In the quantum model, the state of relative motion evolves according to the von Neumann equation:
\begin{align}
i \hbar \partial_t \rho(r,r',t) =& - \frac{\hbar^2}{m} \big( \partial^2_{r} - \partial^2_{r'} \big) \rho(r,r',t) 
\nonumber \\
&+ \big( V(r) - V(r') \big) \rho(r,r',t),
\label{EQ_VN}
\end{align}
where $\rho(r,r',t)$ is the statistical operator in position representation.

To compare classical and quantum dynamics on an equal footing, we employ the Wigner–Weyl transformation. The Wigner function maps quantum operators to phase-space quasi-probability distributions.
For a state $\rho$ this is given by
\begin{align}
W(x,p) = \frac{1}{2 \pi \hbar} \int dy \, e^{-\frac{i}{\hbar} p y } \mel{ x + \frac{y}{2} }{\rho}{ x - \frac{y}{2} } .
\end{align}
Conversely, any classical phase-space distribution $f(x,p)$ can be mapped to a Weyl operator:
\begin{align}
\rho_f(x,x') = \frac{1}{2 \pi \hbar} \int dp \, e^{\frac{i}{\hbar} p (x-x') } f\left(\frac{x+x'}{2},p\right) ,
\end{align}
which need not be positive semi-definite.
There are many possible such mappings~\cite{Walls2008}, but the Wigner function is especially useful because its reconstruction from quadrature measurements parallels the classical reconstruction of $f(x,p)$.

Finally, following Bohm and Hiley~\cite{Bohm1981}, the classical evolution under the Hamiltonian for the relative motion $H_r$,
maps via the Wigner–Weyl transformation to the Weyl-operator evolution equation:
\begin{align}
i \hbar \partial_t \rho_f(r,r',t) = & - \frac{\hbar^2}{m}\left( \partial^2_{r} - \partial^2_{r'} \right) \rho_f(r,r',t)
\nonumber \\
& +  (r-r') \frac{dV}{dr} |_{\frac{r+r'}{2}} \rho_f(r,r',t) .
\label{EQ_BH}
\end{align}

\section{Classical model of experimental signatures of entanglement}
\label{SEC_2ORD}

We are now in a position to explain how classical mechanics can reproduce signatures of entanglement.
Recall that relative displacement $r$ is much smaller than the initial separation $L$,
which justifies expanding the Newtonian potential in powers of $r/L$:
\begin{align}
V(r) \approx V_N(r) 
=  - \frac{1}{4} m \omega^2\sum_{n=0}^N (-1)^n \frac{r^n}{L^{n-2}} ,
\label{EQ_V_N}
\end{align}
where $\omega^2 = 4 G m / L^3$ characterizes the gravitational coupling and $N$ denotes the order of approximation.
For $N=2$, the quantum evolution of the Gaussian state \eqref{EQ_ISTATE} generates entanglement, which has been quantified analytically in Refs.~\cite{Krisnanda2020,Qvarfort2020,Miao2020,Datta2021,Kumar2023}.
Remarkably, the classical dynamics produces exactly the same Weyl operator. Indeed, the von Neumann equation \eqref{EQ_VN} and the Bohm-Hiley equation \eqref{EQ_BH} coincide for any potential that is at most quadratic in $r$.
The only difference between the two dynamical equations comes from the interaction part, but for any quaratic potential 
$V_2(r) = c_0 + c_1 r + c_2 r^2$
we have:
\begin{align}
(r-r') \dv{V_2}{r} |_{\frac{r + r'}{2}} &= c_1 (r - r') + c_2 (r^2 - r'^2) \nonumber \\
&= V_2(r) - V_2(r') .
\end{align}
Accordingly, for initial states with positive Wigner functions, such as in \eqref{EQ_ISTATE}, quantum evolution under a quadratic potential produces a state identical to the Weyl state of the classical dynamics.
This correspondence is well known in quantum optics~\cite{Schleich2001}, and in phase-space treatments of quantum dynamics~\cite{Polkovnikov2010,bartlett_reconstruction_2012}.

How can classical mechanics reproduce quantum-like correlations? The Wigner function of each particle in \eqref{EQ_ISTATE} is Gaussian in both position and momentum, and therefore admits an interpretation as a classical phase-space distribution describing particles with well-defined positions and momenta.
Although the Newtonian interaction is nonlocal in the particle separation, it can be viewed as an effective approximation to an underlying relativistically local theory, in which disturbances propagate at the speed of light~\cite{Caslav_LQG}.
In this respect, the classical dynamics resembles a protocol based on local operations and classical communication (LOCC). Since entanglement is defined as a resource that cannot be increased by LOCC~\cite{Bennett1996}, it is striking that the classical model reproduces signatures ordinarily associated with entanglement.
The resolution lies in the fact that the classical phase-space framework is strictly broader than the quantum one. 
Classical phase space admits arbitrarily sharp distributions, including delta functions that violate the Heisenberg uncertainty relation. 
The Wigner function of the evolved system can be written as a convex combination of products of delta distributions, while at the same time it cannot be decomposed into a convex combination of product distributions that individually satisfy the uncertainty relations.
In a complementary view, the classical phase space emerges as a coarse-grained limit of quantum phase space, with regions of area $\hbar$ identified as single effective points \cite{bibak_classical_2025}. From this perspective, finite-resolution quadrature measurements are unable to resolve the distinction between the classical and quantum descriptions.

\section{Distinguishing between the models}

To distinguish classical from quantum dynamics, one must either start with (local) non-classical states that are Wigner-negative, or increase sensitivity to the third-order coupling in the potential, or go beyond quadrature measurements and other assumptions. We now explore each of these possibilities.

\subsection{Non-classical initial states}

The simplest approach is to prepare the masses in states with a negative Wigner function.
For example, each particle can be initialized in the excited states of a harmonic oscillator.
Using the same excited state for both particles leads to particularly elegant results, but qualitatively similar conclusions hold if only one particle exhibits Wigner negativity.

Quadratic Hamiltonians generate linear symplectic maps in phase space, which preserve the volume of Wigner negativity. As a result, the dynamics remain non-classical at all times. They also generate entanglement, which can be quantified using the covariance matrix, see Appendix~\ref{APP_CV}. For two particles prepared in the $n^\mathrm{th}$ excited state, the covariance matrix $\bm{\sigma}^{(n,n)}(t)$ is related to the covariance matrix when they are prepared in ground states $\bm{\sigma}^{(0,0)}(t)$ via
$\bm{\sigma}^{(n,n)}(t) = (2n+1) \bm{\sigma}^{(0,0)}(t)$.
This leads to a simple scaling of entanglement measured by logarithmic negativity (see Appendix~\ref{APP_LOGN} for explicit derivation):
\begin{align}
\mathcal{E}^{(n,n)}(t) = \max \left[ 0,  \mathcal{E}^{(0,0)}(t) - \log_2(2n+1) \right] .
\end{align}
Since $\mathcal{E}^{(0,0)}(t)$ grows monotonically with time~\cite{Krisnanda2020,Kumar2023}, 
there is always a sufficiently long evolution when the system becomes entangled.

However, it is advantageous to start with an asymmetric situation where one of the masses is in the ground state and the other one in the $n^\mathrm{th}$ excited state.
In such a case, the entanglement accumulates immediately, with the logarithmic negativity at short times scaling as:
\begin{align}
\mathcal{E}^{(0,n)}\qty(t \ll 1/\omega) \approx
\dfrac{2n+1}{8n(n+1) \ln(2)} 
\,
\frac{\omega^4 t^2}{(\hbar/2m\sigma^2)^2} 
.
\end{align}

Notably, the BMV proposal falls into this category~\cite{Bose2017,Marletto2017}. There, each of the two masses is initially prepared in an extended spatial superposition akin to a Schrödinger cat. The Wigner function of such states is known to exhibit significant negativity in the interference region of phase space, which is in-between the displaced coherent states representing the two superposed paths~\cite{Walls2008}. Within the quadratic approximation of the gravitational potential, this negativity is preserved at all times.

\subsection{Sensitivity to third-order coupling}

Alternatively, one may retain Wigner-positive initial states, provided the experiment is sensitive to the third-order term in the expansion \eqref{EQ_V_N}.
We therefore consider both classical phase-space dynamics and quantum Hilbert-space evolution under the same Hamiltonian and identical initial conditions. In this regime, the two descriptions diverge, allowing us to identify operational witnesses of both non-classicality and non-quantumness.

\subsubsection{Quantum dynamics}

\begin{figure}[!t]
\centering
\includegraphics[width=\linewidth]{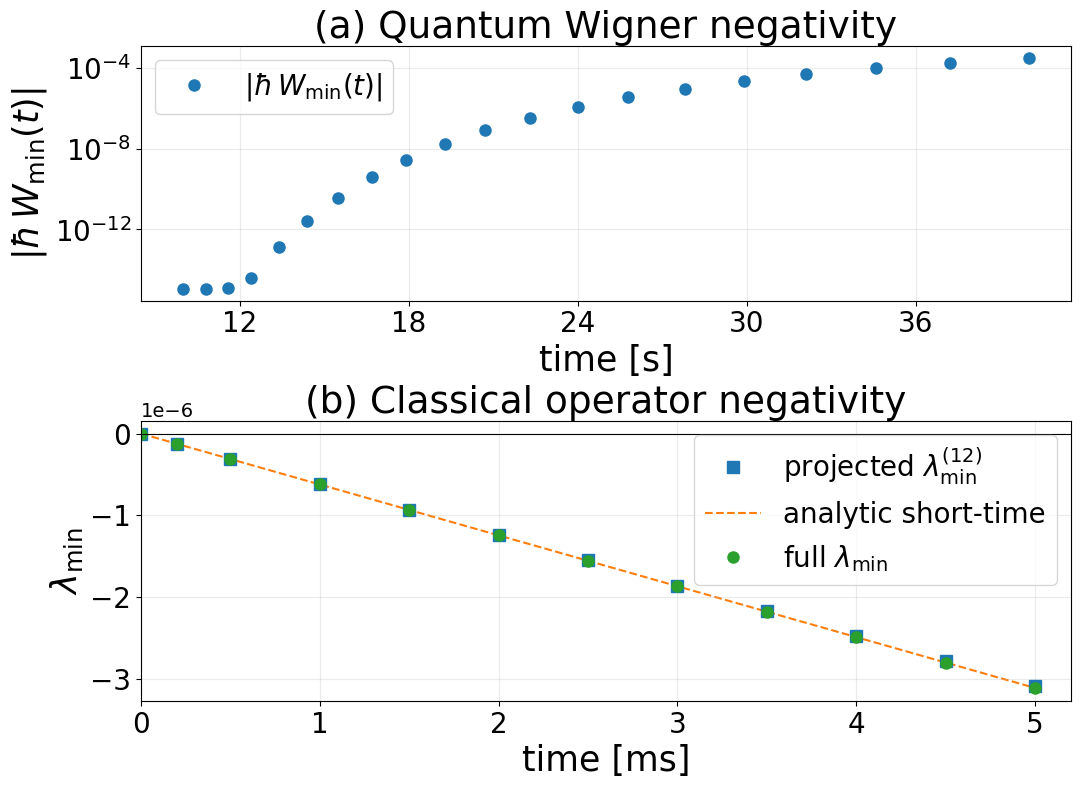}
\caption{Non-classicality of quantum dynamics and non-quantumness of classical dynamics.
All plots are for the Newtonian potential truncated at the cubic term, $V_3(r)$, and the initial state is the product of two identical Gaussians. The markers denote numerically obtained data with parameters from Appendix \ref{APP_WWITNESS}.
(a) Quantum dynamics: Minimum value of the Wigner function $|\min_{r,p}\,\hbar W(r,p,t)|$ over long timescales, showing the emergence of phase-space negativity at the level of $\sim 10^{-4}$. 
(b) Classical dynamics: Smallest eigenvalue of the full Weyl operator $\lambda_\text{min}$ (green circles), and the smallest eigenvalue in the subspace spanned by the first and second excited states at short times $\lambda_\text{min}^{(12)}$ (blue squares). The dashed line is the approximation calculated in \eqref{eq:short_time_lamb}.}
\label{FIG_WIGNEG}
\end{figure}

The non-classicality of quantum dynamics generated by the third-order potential $V_3(r)$ follows directly from standard results. Starting from a pure Gaussian state, a unitary evolution under a non-quadratic Hamiltonian drives the state out of the Gaussian set while preserving purity of the bipartite state~\cite{weedbrook_gaussian_2012}. This departure from Gaussianity can also be seen directly at the level of statistical moments, e.g., from the Moyal equation:
\begin{align}
\label{eq:moyal_3rd}
    \partial_t W
=
-\frac{p}{m}\partial_r W
+ V_3'(r)\partial_p W
-\frac{\hbar^2}{24} V_3'''(r)\partial_p^3 W.
\end{align}
The centered third moment, skewness of momentum, $\mu_{3,p}:=\langle (p-\langle p\rangle)^3\rangle$, for an initially centered Gaussian state, satisfies:
\begin{align}
\left.\frac{d}{dt}\mu_{3,p}\right|_{t=0}
=
\frac{\hbar^2}{4}\,\langle V_3'''(r)\rangle .
\label{EQ_SKEWNESS}
\end{align}
Hence, any nonzero $V_3'''$ immediately generates skewness, while quadratic dynamics preserves vanishing skewness.
By Hudson’s theorem, any pure non-Gaussian state must exhibit Wigner negativity~\cite{hudson_when_1974}. This behavior is illustrated in Fig.~\ref{FIG_WIGNEG}.

A variety of witnesses for Wigner negativity have been proposed~\cite{chabaud_witnessing_2021}.
Here we emphasize that randomized quadrature measurements provide a particularly convenient approach, see Refs.~\cite{vogel_determination_1989, fiurasek_witnessing_2013} for other applications.
A negativity witness is an operator $\hat{\mathcal{N}}$ such that
\begin{align}
\mathcal{N} = \mathrm{tr}(\rho \hat{\mathcal{N}}),
\label{EQ_N}
\end{align}
takes negative values only if the Wigner function $W_{\rho}$, corresponding to the density operator $\rho$, is negative.
A simple construction is to choose $\hat{\mathcal N}$ whose Wigner–Weyl transform $W_\mathcal{N}$ is sharply localized around a phase-space point where $W_{\rho}$ is expected to be most negative. Such an operator cannot correspond to a density matrix, since $\mathrm{tr}(\rho \tilde{\rho}) \ge 0$, for all positive semi-definite $\rho$ and $\tilde{\rho}$.
To connect this witness with experimentally accessible observables, we expand $\hat{\mathcal N}$ in the quadrature basis:
\begin{align}
\hat{\mathcal{N}}  = \iint d\phi \, d x_{\phi} \, \Gamma(\phi, x_{\phi}) | x_{\phi} \rangle \langle x_{\phi} |,
\end{align}
where $| x_{\phi} \rangle$ are eigenstates of the quadrature operator $x_\phi = r \cos(\phi) + p \sin(\phi)$.
This gives
\begin{align}
\mathcal{N} = \iint d\phi \, d x_{\phi} \,  \Gamma(\phi, x_{\phi}) \, \mathrm{tr}(\rho | x_{\phi} \rangle \langle x_{\phi} |).
\label{EQ_GAMMA}
\end{align}
Since the trace yields the probability distribution of quadrature outcomes, $\mathcal{N}$ can be estimated as an average of $\Gamma(\phi, x_{\phi})$ over measurement data. A uniform and random sampling of $\phi$ avoids the need for full state tomography. Moreover, the relative-displacement quadratures can be reconstructed from local quadrature measurements performed on each mass at the same angle $\phi$.

As a concrete example, let $W_\mathcal{N}$ be a Gaussian in phase space, with widths $\Delta_r $ in position and $\Delta_p$ in momentum, satisfying $\Delta_r \Delta_p < \hbar/2$, centered around $( \tilde r_0, \tilde p_0)$. For simplicity we now consider the case $\Delta_r = \Delta_p \equiv \Delta$ in dimensionless quadratures $(\tilde r, \tilde p$) rescaled by the natural frequency $\omega$ (see Appendix \ref{APP_WWITNESS}), leading to the corresponding kernel:
\begin{align}
\label{eq:Gamma_tomo}
\Gamma(y) = \frac{1}{2 \Delta^2} \left[ 1 - \sqrt{\frac{\pi}{2}} \frac{y}{\Delta} e^{- \frac{y^2}{2 \Delta^2}} \mathrm{erfi}\left( \frac{y}{\sqrt{2} \Delta} \right) \right],
\end{align}
where $y = x_{\phi} - x_{\phi}( \tilde r_0, \tilde p_0)$. Averaging this function over randomized quadrature measurements provides a direct witness of Wigner negativity.
Numerical simulation using the parameters of Appendix \ref{APP_WWITNESS} (Fig. 1a) gives $|\hbar W_{\min}| \sim 10^{-4}$ on a timescale of tens of seconds, consistent with the perturbative estimate given there. While this negativity is significant, its detection remains challenging due to the sampling complexity discussed in Appendix \ref{APP_WWITNESS}.

\subsubsection{Classical dynamics}
\label{sec:classical_dyn}
Gravitational entanglement experiments are particularly valuable because the quantum predictions may ultimately fail to match observations. We therefore analyze the corresponding classical dynamics and identify signatures that distinguish it from the quantum model.

We decompose the Bohm–Hiley equation \eqref{EQ_BH} into a quadratic part generated by $H_2 = - \frac{\hbar^2}{m} \partial^2_r - \frac{1}{4} m \omega^2 r^2$, and a cubic correction:
\begin{align}
\mathcal{V}_3(r,r') = (r-r') \frac{3 m \omega^2}{4 L} \left( \frac{r + r'}{2}\right)^2.
\end{align}
The initial Gaussian state of the relative displacement corresponds to the ground state $\chi_0(r)$ of a harmonic oscillator. We show that the Weyl operator $\rho_f(r,r',t)$ immediately develops a negative eigenvalue within the subspace spanned by $\{ \chi_1(r), \chi_2(r) \}$, i.e., the first and second excited states of the oscillator.
Initially, there is no population in this subspace, and the quadratic dynamics alone does not generate it. This follows from the commutator structure of the von Neumann evolution, which ensures $\langle 2 | [H_2, \rho_0] | 1 \rangle = 0$, given $\rho_0 = | 0 \rangle \langle 0 |$, where we used the standard ket notation $| n \rangle$ for the position space eigenfunctions $\chi_n(r)$.
At short times, the Hamiltonian is irrelevant, and the  vanishing overlap is solely due to vanishing inner products with the initial state.
In this sector, the dynamics is governed entirely by the cubic term:
\begin{align}
\dot \rho_{jk} = - \frac{i}{\hbar} \iint dr dr' \chi_j^*(r) \mathcal{V}_3(r,r') \chi_0(r) \chi_0^*(r') \chi_k(r').
\end{align}
Evaluating this expression yields a nonzero contribution for $(j,k)=(1,2)$ or $(2,1)$, leading to a negative eigenvalue at short times:
\begin{align}
\lambda\qty(t \ll 1/\omega) \approx - \frac{3 m \omega^2 \sigma^3}{4 \hbar L}  t.
\label{eq:short_time_lamb}
\end{align}

This immediately suggests a witness for non-quantumness. Using the same randomized quadrature protocol, one can evaluate \eqref{EQ_N} with $\hat{\mathcal N}$ chosen as the projector onto the state $\frac{1}{\sqrt{2}}(|1 \rangle + i | 2 \rangle)$, which detects the negative eigenvalue, see Fig.~\ref{FIG_WIGNEG}b.

Note a duality between the two types of witnesses.
To detect Wigner negativity, $\rho$ is a valid quantum state, while $\hat{\mathcal N}$ corresponds to a non-physical operator (violating the uncertainty principle). Conversely, to detect non-quantumness, $\rho_f$ is not a physical quantum state (it develops negative eigenvalues), while $\hat{\mathcal N}$ is a projector onto a legitimate quantum state.

\subsubsection{Witnessing third-order coupling}

As demonstrated above, experimental sensitivity to the cubic contribution in the expansion~(\ref{EQ_V_N}) of the gravitational potential enables one to distinguish between the quantum and classical evolution models.
This sensitivity can be identified through the emergence of non-zero skewness, see \eqref{EQ_SKEWNESS}, but it may also be witnessed using only first moments.
An example is provided by the time evolution of the quantity
\begin{align}
\mathcal{C} = \frac{1}{m} \ev{p}^2 - \frac{1}{4} m \omega^2 \qty( \ev{r} - \frac{L}{2} )^2  .
\end{align}
To make the role of the cubic correction explicit, we introduce a parameter $\theta$ into the truncated gravitational potential
\begin{align}
V_3(r) = - \frac{1}{4} m\omega^2 \qty( L^2 - Lr + r^2 - \theta \frac{r^3}{L} )  .
\end{align}
The Ehrenfest’s theorem then implies
\begin{align}
\dot{\mathcal{C}} = - \theta \frac{3\omega^2}{2L} \ev{r^2} \ev{p}  .
\end{align}
Accordingly, for a purely quadratic potential ($\theta = 0$), the quantity $\mathcal{C}$ remains constant in time. 
Any observed time dependence therefore directly signals sensitivity to the cubic term in the gravitational interaction. 
Notably, the dependence on $\ev{p}$ implies that the witness becomes non-trivial  only when the masses possess large relative momentum, e.g., they are pushed towards each other.

An even simpler witness is the cross-axis correlation generated from two-dimensional Gaussian initial states.
Let the symmetry axis be $\vec{L} = L \hat e_x$, and write the displacement vectors of the two masses from their initial mean positions as $\vec{r}_j = x_j \hat e_x + y_j \hat e_y$.
For the two-dimensional Newtonian potential
$ V(\vec{r}_1,\vec{r}_2) 
= - Gm^2 / |\vec{L}+\vec{r}_2-\vec{r}_1| $,
a multipole expansion yields the leading correlation-generating terms:
\begin{align}
V
\equiv&  \frac{1}{2} m\omega^2 \qty( x_1x_2 
- \frac{1}{2} y_1y_2 )
\\
&+ \frac{3}{4L} m\omega^2 \qty(
(x_2-x_1)y_1y_2 + \frac{1}{2}(x_1y_2^2 - x_2y_1^2 ) )
.
\nonumber
\end{align}
Note that correlations between orthogonal directions of the two particles (cross-axis correlations, e.g., $\langle x_1 y_2^2 \rangle$) can only arise from cubic or higher-order terms.
Accordingly, the observation of any non-vanishing form of correlation along orthogonal directions directly indicates sensitivity to third-order couplings.

\section{Discussion}

The methods discussed so far, including those implemented in current experiments, rely on quadrature measurements of the motional degrees of freedom. In optomechanical setups, the motion of the masses is first mapped onto the state of an optical field and subsequently inferred through homodyne detection. Such measurements possess direct analogues in both classical and quantum descriptions.
This naturally raises the question about implications of employing observables with non-classical Wigner-Weyl representation.

A well-known example is displaced parity, which was shown to reveal Bell inequality violations even for appropriate two-mode Gaussian states~\cite{Banaszek1998}. Such measurements would be particularly rewarding in the gravitational context because they can rule out the LOCC-like classical models discussed in Sec.~\ref{SEC_2ORD}.
These models are deterministic and local, with the positions and momenta of the particles determining the outcomes of all other observables.
However, displaced parity has no counterpart in classical mechanics. The significance of Bell violation is that no extension of the classical model can consistently incorporate such an observable without abandoning at least one assumption underlying Bell’s theorem.
 Moreover, classical mechanics satisfies both setting and outcome independence and therefore any model in which observables are determined by classical positions and momenta can be ruled out even through measurements on nearby masses. This conclusion is reached while starting from Gaussian states and considering only quadratic gravitational dynamics.

In optical systems, photon counting provides a prominent example of a non-quadrature measurement, directly probing the discreteness of the electromagnetic field.
Observations such as sub-Poissonian photon statistics~\cite{Kimble1977} and the Hong-Ou-Mandel dip~\cite{Hong1987} offered a fresh perspective on the non-classical nature of light.
Although the detection of individual gravitons may be fundamentally impossible, see e.g.~\cite{Dyson2013}, it remains an interesting open research direction to identify measurement schemes capable of probing discreteness effects associated with larger numbers of gravitons.

Finally, recall that the classical explanation of entanglement signatures in the Gaussian regime relies on modeling the masses as point particles, i.e., objects that do not obey the uncertainty relations.
Another possible route toward excluding such classical descriptions is therefore to obtain independent evidence that the individual masses themselves cannot be classical.
We have already covered preparations involving states with Wigner negativity, but it would be experimentally relevant to determine whether additional signatures exist that, when combined with strictly Wigner-positive gravitational evolution and simple measurements, would nevertheless rule out the classical models.

\section{Conclusions}

Classical mechanics can reproduce signatures of gravitational entanglement when experiments are restricted to Wigner-positive initial states, second-order expansion of the gravitational potential, and quadrature measurements.
This regime coincides with the standard operating conditions of present-day optomechanical experiments, highlighting the importance of moving beyond the current experimental paradigm.
As shown, the most direct route towards demonstrating genuinely quantum behaviour is to prepare individual masses in non-classical states.
Alternatively, sensitivity to third-order corrections in the gravitational potential leads to a divergence between classical and quantum predictions, even for Wigner-positive states.
We provided explicit witnesses demonstrating this divergence that can reveal non-classicality as well as non-quantumness.
These should be useful guides for future experiments in the field.

\section{Acknowledgments}

The authors warmly thank Flavio Del Santo, Markus Aspelmeyer, and Časlav Brukner, for stimulating discussions. 
This work is supported by the National Science Centre (NCN, Poland) within the OPUS project (Grant No. 2024/53/B/ST2/04103).
A.K. is partially/fully supported by the Helen Diller Quantum Center at the Technion. 
This research was funded in whole, or in part, by the Austrian Science Fund (FWF) [10.55776/F71], [10.55776/P36994] and [10.55776/COE1] and the European Union– NextGenerationEU. For open access purposes, the author(s) has applied a CC BY public copyright license to any author accepted manuscript version arising from this submission.

\appendix

\section{Covariance matrix dynamics}
\label{APP_CV}

Consider laboratory (LAB) phase space variables $\mathcal{X} = (x_1, p_1, x_2, p_2)^T$ and the corresponding vector after their change to the COM frame  $\mathcal{R} = (R,P,r,p)^T$.
They are related through $\mathcal{X} = M \mathcal{R}$, where
\begin{align}
M = \mqty[
1 & 0 & -1/2 & 0 \\
0 & 1/2 & 0 & -1 \\
1 & 0 & 1/2 & 0 \\
0 & 1/2 & 0 & 1
] .
\end{align}
The Hamiltonian in the COM frame, truncated at the second-order term, is given by:
\begin{align}
H &= \frac{P^2}{4 m} + \frac{p^2}{m} - \frac{1}{4} m \omega^2 (L^2 - L r + r^2) 
\nonumber \\
&\equiv \frac{P^2}{4 m} + \frac{p^2}{m} - \frac{1}{4} m \omega^2  r^2 ,
\end{align}
where we ignored the constant term as it accumulates only a global phase. The linear term has also been dropped as it produces a uniform position-independent force that does not contribute to the covariance matrix.
For the defined quadrature vectors, the covariance matrix is a collection of all (centered) second moments:
\begin{align}
\bm{\sigma}_{jk} &= \frac{1}{2} \langle \mathcal{X}_j \mathcal{X}_k + \mathcal{X}_k \mathcal{X}_j \rangle - \langle \mathcal{X}_j \rangle \langle \mathcal{X}_k \rangle ,
\nonumber\\
\bm{\varsigma}_{jk} &= \frac{1}{2} \langle \mathcal{R}_j \mathcal{R}_k + \mathcal{R}_k \mathcal{R}_j \rangle - \langle \mathcal{R}_j \rangle \langle \mathcal{R}_k \rangle  ,
\end{align}
with both these representations related to each other through $\bm{\sigma}(t) = M \bm{\varsigma}(t) M^T $.
The covariance matrix $\bm{\varsigma}$ evolves according to a linear symplectic map:
$ \bm{\varsigma}(t) = S(t) \bm{\varsigma}(0) S(t)^T $,
where the matrix $S(t)$ is determined from the exponentiation of the drift matrix in the Ehrenfest's equation of motion for the first moments: $\ev{\dot \xi} = A \ev{\xi} $.
In our case we have the drift matrix and the corresponding exponentiation as:
\begin{align}
A &= 
\mqty[  0 & \frac{1}{2m} \\[0.5ex]
			0 & 0 ] 
\oplus
\mqty[ 0 & \frac{2}{m} \\[0.5ex]
	   \frac{m \omega^2}{2} & 0 
] ,
\\[1ex]
S(t) &= 
	\mqty[ 1 & \frac{t}{2m} \\[0.5ex]
	0 & 1 	] 
\oplus
\mqty[ \cosh(\omega t) & \frac{2}{m \omega} \sinh(\omega t) \\[0.5ex]
	\frac{m \omega}{2} \sinh(\omega t) & \cosh(\omega t) ] .
\nonumber
\end{align}

For the two masses prepared in ground states the initial quantum state separates and the covariance matrix is block diagonal.
However, if at least one particle is prepared in an excited state, the mapping leads to an entangled initial state in the COM frame.
Assuming that the masses start from the Fock states $n_1$ and $n_2$, the initial covariance matrix in LAB frame is diagonal with:
$  \bm{\sigma}(0) =
  \mathrm{diag} 
  \big[ 
  (2n_1+1)  \sigma^2, \, (2n_1+1) \hbar^2 / 4\sigma^2, \,
 (2n_2+1) \sigma^2, \, (2n_2+1) \hbar^2 / 4\sigma^2
 \big]  $. 
In the special symmetric case, $n_1=n_2=n$, the LAB frame covariance matrix $\bm{\sigma}$ is simply $2n+1$ times the covariance matrix when starting from the ground states, which was obtained in exact closed form in Ref.~\cite{Kumar2023}.

\section{Logarithmic negativity}
\label{APP_LOGN}

Consider a bipartite system, where the $4 \times 4$ covariance matrix could be written in the block form:
\begin{align}
\bm{\sigma} = \mqty[
\bm{\alpha} & \bm{\gamma} \\
\bm{\gamma}^T & \bm{\beta}
] ,
\end{align}
where $\bm\alpha$ and $\bm\beta$ encode the local information of the two modes, and $\bm{\gamma}$ contains their intermodal correlations.
The negativity of the partially transposed density matrix is a necessary and sufficient condition for entanglement in two--mode Gaussian states~\cite{simon_peres-horodecki_2000, duan_inseparability_2000}. 
Such a partial transposition of the density matrix is equivalent to a transformation of the covariance matrix to $\tilde{\bm{\sigma}}$, which differs from $\bm{\sigma}$ only through a sign-flip of $\det (\bm{\gamma})$~\cite{adesso_entanglement_2007, serafini_quantum_2023}.
The symplectic eigenvalues of the partially transposed system are~\cite{adesso_extremal_2004,vidal_computable_2002}:
\begin{align}
\tilde{\nu}_{\pm} (\bm{\sigma}) = 
\frac{1}{\sqrt{2}} 
\sqrt{
\tilde{\Sigma}(\bm{\sigma}) \pm \sqrt{\tilde{\Sigma}^2(\bm{\sigma}) - 4 \det( \bm{\sigma} ) }
} ,
\end{align}
where $\tilde{\Sigma}(\bm{\sigma}) = \det (\bm{\alpha}) + \det ( \bm{\beta}) - 2 \det ( \bm{\gamma} )$.
Entanglement is quantified by the minimum symplectic eigenvalue via logarithmic negativity:
\begin{align}
\mathcal{E}(\bm{\sigma}) = \max \qty[ 0, \ -\log_2\qty(  \frac{2 \tilde\nu_-(\bm{\sigma)}}{\hbar} ) ].
\label{eq:E_from_covmat}
\end{align}

\section{Wigner negativity witness}
\label{APP_WWITNESS}

In the following we work with dimensionless phase space parameters defined as: 
\begin{align}
\tilde r =  \frac{r}{\sqrt{\hbar/m\omega}} ,
\quad
\tilde p = 
 \frac{p}{\sqrt{m\hbar\omega}} .
\end{align}
and satisfying $[\tilde r, \tilde p] = i$.
To witness the negativity of the Wigner function at a phase space point $(\tilde r_0, \tilde p_0)$, we start with the windowed negativity witness
\begin{align}
\label{eq:app_neg}
    \mathcal{N}(\tilde r_0, \tilde p_0) := \iint d\tilde r \, d\tilde p \, G_\Delta(\tilde r, \tilde p)\, W_t(\tilde r,\tilde p),
\end{align}
where $W_t$ is the relative-mode Wigner function at time $t$, and 
\begin{align}
    G_\Delta(\tilde r, \tilde p) = \frac{1}{2 \pi \Delta^2} 
    \exp\!\left[ - \frac{(\tilde r-\tilde r_0)^2 }{2\Delta^2} - \frac{(\tilde p-\tilde p_0)^2}{2\Delta^2} \right] ,
\end{align}
is an isotropic Gaussian window of width $\Delta < 1/\sqrt{2}$, the bound enforced by the requirement that $G_\Delta$ does not correspond to a valid quantum state. Since $G_\Delta \geq 0$, any negative value of $\mathcal{N}$ implies that $W_t$ is negative somewhere in the support of the window. To express $\mathcal{N}$ in terms of quadrature marginals, we use the inverse Radon representation \cite{Mauro_D_Ariano_2003} of $W_t$, substitute it into \eqref{eq:app_neg}, and carry out the resulting integrations explicitly. This gives
\begin{align}
    \mathcal{N}(\tilde r_0, \tilde p_0) = \int_0^\pi \frac{d\phi}{\pi} \int_{-\infty}^\infty dx \, p_t(x|\phi) \, \Gamma_\Delta(y),
\end{align}
where $p_t(x|\phi)$ is the probability density for the quadrature $x_\phi = \tilde r \cos\phi + \tilde p \sin\phi$, $y = x - x_\phi(\tilde r_0, \tilde p_0)$, and the pattern function \cite{lvovsky_continuous-variable_2009} reads
\begin{align}
\label{eq:app_gamma}
    \Gamma_\Delta(y) = \frac{1}{2 \Delta^2} \!\left[ 1 - \sqrt{\frac{\pi}{2}}\frac{y}{\Delta} \, e^{-\frac{y^2}{2\Delta^2}}\,\text{erfi}\!\left(\frac{y}{\sqrt{2}\Delta}\right) \right],
\end{align}
as quoted in the main text. The relative-mode quadrature statistics are obtained from same-angle local quadrature measurements on the two masses, so $\mathcal{N}$ is directly estimable from data without full state tomography.

\textit{Order-of-magnitude estimate of Wigner negativity.}
The existence of Wigner negativity under the cubic quantum dynamics is argued in the main text from the generation of non-Gaussianity together with Hudson's theorem. Here we use the first-order Moyal correction as an order-of-magnitude estimate for the location and depth of the negative feature. The cubic part of the potential,
\begin{align}
    \mathcal{V}_3(r) = \frac{m\omega^2}{4L}  r^3 ,
\end{align}
has a constant third derivative, and therefore contributes to the Moyal evolution only through the term
\begin{equation}
\left.\partial_t W\right|_{\mathrm{Moyal}}
=
-\frac{\hbar^2 m\omega^2}{16L}\,\partial_p^3 W .
\end{equation}
For short times, and neglecting the comparatively small phase-space rotation generated by the quadratic part, this gives the perturbative estimate
\begin{align}
    \delta W(r,p,t)
    \approx
    -\frac{\hbar^2 m\omega^2 t}{16L}\,\partial_p^3 W_0(r,p).
\end{align}
For the initial Gaussian state with momentum standard deviation $\sigma_p$,
\begin{align}
    \partial_p^3 W_0
    =
    \left(
        \frac{3p}{\sigma_p^4}
        -
        \frac{p^3}{\sigma_p^6}
    \right)W_0 ,
\end{align}
so that along the $r=0$ axis
\begin{align}
\label{eq:app_Wt_pert}
\begin{split}
    \frac{W_t(0,p)}{W_0(0,p)}
    \approx&
    1
    +
    \epsilon
    \left[
        \left(\frac{p}{\sigma_p}\right)^3
        -
        3\frac{p}{\sigma_p}
    \right],
    \\
    \epsilon
    :&=
    \frac{\hbar^2 m\omega^2 t}{16L\sigma_p^3}.
\end{split}
\end{align}
For the parameters used below, $\epsilon \approx 5\times 10^{-2}$. The correction is therefore small near the bulk of the Gaussian, and a sign change can only occur where the cubic polynomial overcomes the leading constant. Parametrically, this happens when
\begin{align}
    \epsilon \left|\frac{p}{\sigma_p}\right|^3 \sim 1.
\end{align}
For $\epsilon \sim 5\times 10^{-2}$ this places the expected negative feature around the three-sigma momentum tail of the initial Gaussian, well within its support.

Minimizing the first-order expression $W_0+\delta W$ over $p$ gives the estimate
\begin{align}
    \tilde p_0 \approx -0.42,
    \qquad
    W_{\rm tail}
    \sim
    -10^{-4}.
\end{align}
This number should be interpreted only as the natural order of magnitude of the negative feature, not as a controlled prediction of the exact minimum, as the sign change occurs where the relative correction $\delta W/W_0$ is of order unity, so higher-order corrections affect the precise location and depth. Numerical simulation of the full nonlinear Moyal evolution at the parameters below confirms the order of magnitude, with the actual minimum sitting at a value of approximately $-2\times 10^{-4}$ and shifted slightly off the $r=0$ axis by the classical drift induced by $V_3'(r)$.

The same estimate can be applied to the Gaussian-window witness. Convolving the perturbative expression with $G_\Delta$ replaces the momentum width by the broadened effective width
\begin{align}
    \sigma_{p,\rm eff}^2
    =
    \sigma_p^2+\Delta_p^2,
\end{align}
where $\Delta_p$ is the momentum width corresponding to the dimensionless window size $\Delta$. This gives the approximate scaling
\begin{align}
\label{eq:app_N_pert}
\begin{split}
    \mathcal{N}(0,p_0;\Delta)
    \approx
    \left[
        1
        +
        \frac{\hbar^2 m\omega^2 t}{16L\sigma_{p,\rm eff}^3}
        \left(
            \frac{p_0^3}{\sigma_{p,\rm eff}^3}
            -
            3\frac{p_0}{\sigma_{p,\rm eff}}
        \right)
    \right]\\ \times
    (G_\Delta * W_0)(0,p_0).
\end{split}
\end{align}
Thus the witness is controlled by the same competition between the cubic polynomial and the Gaussian tail. Within this order-of-magnitude treatment, the windowed witness remains negative only for sufficiently narrow windows. For the parameters below, the perturbative estimate gives a critical scale of order
\begin{align}
    \Delta_* \sim 4\times 10^{-2},
    \label{EQ_DENTAS}
\end{align}
well below the admissibility threshold $\Delta<1/\sqrt{2}$. Around this value, the magnitude of the optimized witness is estimated to be of the same order as the tail value,
\begin{align}
    |\mathcal{N}_{\rm opt}|
    \sim
    10^{-4}.
    \label{EQ_N_EST}
\end{align}

\textit{Sample complexity.}
The number of randomized quadrature samples required to resolve the negativity of the witness value at one-standard-deviation confidence scales as
\begin{align}
\label{eq:app_M}
    \mathcal{M}
    \sim
    \frac{\mathrm{Var}(\Gamma_\Delta)}
    {|\mathcal{N}|^2}.
\end{align}
The pattern function obeys
\begin{align}
    \Gamma_\Delta(y)
    =
    \Delta^{-2}f(y/\Delta)
\end{align}
for a fixed dimensionless profile $f$. Therefore a change of variables in $\langle \Gamma_\Delta^2\rangle$ gives
\begin{align}
\label{eq:app_var}
\begin{split}
    \mathrm{Var}(\Gamma_\Delta)
    &\approx
    \frac{C_\Gamma}{\Delta^3},
    \\
    C_\Gamma
    &=
    \left\langle
        p_t\!\left(x_\phi(\tilde r_0,\tilde p_0)\middle|\phi\right)
    \right\rangle_\phi
    \int du\, f(u)^2 ,
\end{split}
\end{align}
where the angular average is taken over uniformly sampled $\phi\in[0,\pi)$. For the parameters considered here, the prefactor is of order
\begin{align}
    C_\Gamma \sim 10^{-1}.
\end{align}
Using the perturbative estimates (\ref{EQ_DENTAS}) and (\ref{EQ_N_EST})
one obtains
\begin{align}
    \mathrm{Var}(\Gamma_{\Delta_*})
    \sim
    \frac{10^{-1}}{(4\times10^{-2})^3}
    \sim
    10^{3} ,
\end{align}
and consequently
\begin{align}
    \mathcal{M}_{\rm opt}
    \sim
    \frac{10^{3}}{(10^{-4})^2}
    \sim
    10^{11}.
\end{align}
For the representative and optimistic parameters we chose:
$m = 0.5$ pg and $L = 470$ nm, which implies the entangling frequency $\omega \approx 2\pi\times 1.81\times10^{-4}$ Hz. The initial wavepacket width is assumed to be $\sigma = 40$ nm, and the final experimental time as $t = 40$ s.
The randomized-quadrature witness therefore requires a large but not prohibitive number of samples. Again, this conclusion should be read as an order-of-magnitude estimate rather than a sharp statistical bound, because the witness value itself was estimated from the perturbative tail and the actual minimum of $W_t$ is shifted slightly by the classical drift in $V_3$.

The scaling, however, shows how to push further into a tractable regime. From Eq.~\eqref{eq:app_Wt_pert}, the perturbation strength increases with larger mass, smaller separation, longer interaction time, and broader initial position width. Substituting $\omega^2 = 4Gm/L^3$ and $\sigma_p = \hbar/(2\sigma_r)$ gives the explicit scaling
\begin{align}
    \epsilon
    \propto
    \frac{m^2 \sigma_r^3 t}{\hbar L^4} \sqrt{G}.
\end{align}
Increasing this parameter moves the negative feature out of the Gaussian tail and reduces the required sample complexity.

\bibliographystyle{quantum}
\bibliography{main, references}

\end{document}